\documentclass[english,aps, twocolumn, pre,superscriptaddress, notitlepage]{revtex4-1}
\usepackage[T1]{fontenc}
\usepackage[latin9]{inputenc}
\usepackage{color}
\usepackage{array}
\usepackage{textcomp}
\usepackage{bm}
\usepackage{amsmath}
\usepackage{amssymb}
\usepackage{graphicx}

\makeatletter

\providecommand{\tabularnewline}{\\}

\usepackage[colorlinks=true,linktocpage=true,linkcolor=MidnightBlue,urlcolor=MidnightBlue,citecolor=MidnightBlue,anchorcolor=MidnightBlue]{hyperref}
\usepackage[dvipsnames]{xcolor}

\makeatother

\usepackage{babel}
\begin{document}

\title{Fast Bayesian inference of the multivariate Ornstein-Uhlenbeck process}

\author{Rajesh Singh }
\email{rs2004@cam.ac.uk}

\affiliation{DAMTP, Centre for Mathematical Sciences, University of Cambridge,
Wilberforce Road, Cambridge CB3 0WA, UK}

\author{Dipanjan Ghosh}

\affiliation{Department of Chemical Engineering, Jadavpur University, Kolkata
700032, India}

\author{R. Adhikari}
\email{rjoy@imsc.res.in, ra413@cam.ac.uk}

\affiliation{The Institute of Mathematical Sciences-HBNI, CIT Campus, Taramani,
Chennai 600113, India}

\affiliation{DAMTP, Centre for Mathematical Sciences, University of Cambridge,
Wilberforce Road, Cambridge CB3 0WA, UK}
\begin{abstract}
The multivariate Ornstein-Uhlenbeck process is used in many branches
of science and engineering to describe the regression of a system
to its stationary mean. Here we present an $O(N)$ Bayesian method
to estimate the drift and diffusion matrices of the process from $N$
discrete observations of a sample path. We use exact likelihoods,
expressed in terms of four sufficient statistic matrices, to derive
explicit maximum a posteriori parameter estimates and their standard
errors. We apply the method to the Brownian harmonic oscillator, a
bivariate Ornstein-Uhlenbeck process, to jointly estimate its mass,
damping, and stiffness and to provide Bayesian estimates of the correlation
functions and power spectral densities. We present a Bayesian model
comparison procedure, embodying Ockham's razor, to guide a data-driven
choice between the Kramers and Smoluchowski limits of the oscillator.
These provide novel methods of analyzing the inertial motion of colloidal
particles in optical traps.\\\\DOI: \href{http://dx.doi.org/10.1103/PhysRevE.98.012136}{10.1103/PhysRevE.98.012136} 
\end{abstract}
\maketitle

\section{Introduction}

The multivariate Ornstein-Uhlenbeck process is widely used in many
branches of science and engineering to describe the regression of
a system to its stationary mean. Its importance arises from the fact
that it is the only continuous stochastic process that is simultaneously
stationary, Gaussian and Markovian \cite{doob1942brownian,doob1953stochastic}.
Therefore, the process is fully characterized by its stationary and
conditional distributions, each of which is a multivariate Gaussian,
with mean and variance determined by the regression rates and diffusion
coefficients of the process \cite{van1992stochastic,gardiner1985handbook}.
This simplicity implies that the likelihood associated with discretely
observed sample paths can be calculated explicitly and exactly in
terms of the process parameters and that the posterior distributions
of the latter can be obtained, without approximation, from discrete
observations. 

Here we present the details of such an exact Bayesian estimation method
for the parameters of the $M$-dimensional Ornstein-Uhlenbeck process.
We show that the likelihood can be expressed in terms of four matrices
whose elements are the self- and mutual-correlations of the stochastic
variables at equal times and one observation time apart. We derive
the maximum a posteriori (MAP) estimates and the error bars of the
parameters in terms of these four sufficient statistics. We address
the problem of model selection, i. e., of choosing the Ornstein-Uhlenbeck
model with the least number of parameters that best explains the data,
within the framework of a Bayesian model comparison. 

The principal computational cost of our method is in evaluating the
four sufficient statistic matrices. For a time series of $N$ discrete
observations, only $O(N)$ operations are required to compute the
necessary self- and mutual-correlations. Further, additional $N^{\prime}$
observations can be incorporated incrementally and the sufficient
statistics recomputed at a cost of $O(N^{\prime})$, making our algorithm
suitable for ``online'' estimation. Together with the consistency
and optimality of Bayesian estimators, this yields a fast and accurate
parameter estimation method for the multivariate Ornstein-Uhlenbeck
process. 

We apply our general results to the Brownian harmonic oscillator,
a bivariate Ornstein-Uhlenbeck process that is ubiquitous in the physical
sciences. This provides a Bayesian method for jointly estimating the
mass, friction, and spring constant of the oscillator and new ways
of estimating its correlation function and power spectral density.
The results are pertinent for the analysis of time-series data of
inertial motion of colloidal particles in optical traps and for selecting
between the underdamped and overdamped models of oscillator motion,
when the parameters are not known \emph{apriori. }

We make a few remarks on how this work relates to the preceding literature.
Bayesian inference for general Gaussian processes, of which the Ornstein-Uhlenbeck
process is a special case, has been an area of extensive research
in the past few decades \cite{mackay2003information,rasmussen2006gaussian,lindgren2011explicit}.
For $N$ observations of the process, inference requires the inversion
of an $N\times N$ symmetric covariance matrix whose computational
cost, using a direct method, is $O(N^{3})$. This makes inference
superlinear in $N$ and for large $N$ approximate methods become
necessary \cite{smola2000sparse,williams2001using,csato2002sparse,quinonero2005unifying}.
Imposing the Markov property on a general Gaussian process reduces
the covariance matrix to symmetric tridiagonal form and, imposing
stationarity further simplifies it to symmetric, tridiagonal and Toeplitz
form. In this limit, the inverse can be obtained analytically in terms
of sufficient statistics, and the only cost is in computing the latter,
which is proportional to the number of data points $N$. This special
feature of the stationary Gauss-Markov processes, which allows for
fast, yet exact, inference, appears not to have been exploited earlier. 

The remainder of the paper is organized as follows. In section \ref{sec:mvoup},
we briefly review key properties of the Ornstein-Uhlenbeck process,
and we make use of them in section \ref{sec:Bayesian-inference} to
obtain MAP estimates for the parameters and the model odds. In section
\ref{sec:Path-sampling}, we present an exact path-sampling algorithm,
which is used in \ref{sec:Brownian-harmonic-oscillator} to generate
sample paths of the Brownian harmonic oscillator and to validate the
results of section \ref{sec:Bayesian-inference}. We conclude in section
\ref{sec:conclusion} with a discussion on further applications. 

\section{Multivariate Ornstein-Uhlenbeck Process\label{sec:mvoup}}

The multivariate Ornstein-Uhlenbeck is defined by the It$\bar{\text{o}}$
stochastic differential equation \cite{gardiner1985handbook}
\begin{equation}
dx_{i}=-\lambda_{ij}x_{j}dt+\sigma{}_{ij}dW_{j},\label{eq:mvou}
\end{equation}
where $-\lambda_{ij}$ is a stable matrix of mean regression rates,
$\sigma{}_{ij}$ is the volatility matrix, $W_{i}(t)$ are Wiener
processes, and $i,j=1,\ldots,M$. We denote $(x_{1},\ldots,x_{M})^{\text{tr}}$
by the vector $\boldsymbol{x}$ and $\lambda_{ij}$ by the matrix
$\boldsymbol{\lambda}$, with similar bold-face notation for other
vectors and matrices, when convenient. The volatility matrix is related
to $\boldsymbol{D}$, a symmetric positive-semi-definite matrix of
diffusion coefficients, as $\boldsymbol{\sigma}\boldsymbol{\sigma}^{\text{tr}}\equiv2\boldsymbol{D}$.

The probability density of a displacement from $\boldsymbol{x}$ at
time $t$ to $\boldsymbol{x}^{\prime}$ at time $t^{\prime}$, $P_{1|1}$$(\boldsymbol{x}^{\prime},t^{\prime}|\boldsymbol{x},t$),
obeys the Fokker-Planck equation $\partial_{t}P_{1|1}=\mathcal{L}P_{1|1}$,
where the Fokker-Planck operator is
\begin{equation}
\mathcal{L}(\boldsymbol{x})=\frac{\partial}{\partial x_{i}}\lambda_{ij}x_{j}+\frac{1}{2}\frac{\partial^{2}}{\partial x_{i}\partial x_{j}}(\boldsymbol{\sigma}\boldsymbol{\sigma}^{\text{tr}})_{ij}.
\end{equation}
The solution is a multivariate normal distribution
\begin{equation}
\mathbf{\boldsymbol{x}}^{\prime},t^{\prime}|\boldsymbol{x},\thinspace t\sim\mathcal{N}\left(\boldsymbol{\mu}\,,\boldsymbol{\Sigma}\right),\label{eq:conditional-probability}
\end{equation}
where 
\begin{equation}
\boldsymbol{\mu}=\boldsymbol{\Lambda}\boldsymbol{x},\quad\boldsymbol{\Sigma}=\boldsymbol{c}-\boldsymbol{\Lambda}\boldsymbol{c}\,\boldsymbol{\Lambda}^{\text{tr}},\quad\boldsymbol{\Lambda}=e^{-\boldsymbol{\lambda}|\Delta t|}
\end{equation}
and $\Delta t=t'-t$. This solution is exact and holds for arbitrary
values of $\Delta t$. The stationary distribution $P_{1}(\boldsymbol{x})$
obeys the steady-state Fokker-Planck equation $\mathcal{L}P_{1}=0$,
and the solution is, again, a normal distribution,
\begin{equation}
\boldsymbol{x}\sim\mathcal{N}\left(\boldsymbol{0},\boldsymbol{c}\right).
\end{equation}
Then, $\boldsymbol{c}=\langle\boldsymbol{x}\boldsymbol{x}^{\text{tr}}\rangle$
can be identified as the matrix of covariances in the stationary state,
and $\mathcal{L}P_{1}=0$ implies that the matrices $\boldsymbol{\lambda}$,
$\boldsymbol{c}$, and $\boldsymbol{\sigma}$ are not all independent
but are related by the stationarity condition
\begin{equation}
\boldsymbol{\lambda c}+(\boldsymbol{\lambda c})^{\text{tr}}=\boldsymbol{\sigma}\boldsymbol{\sigma}^{\text{tr}}.\label{eq:stationarity}
\end{equation}
This is a Lyapunov matrix equation for $\boldsymbol{c},$ given $\boldsymbol{\lambda}$
and $\boldsymbol{\sigma}$. Solutions are considerably simplified
when the Fokker-Planck operator obeys detailed balance $\mathcal{L}(\boldsymbol{x)}P_{1}(\boldsymbol{x})=P_{1}(\boldsymbol{\epsilon x})\mathcal{L}^{\dagger}(\boldsymbol{\epsilon x})$,
where $\mathcal{L^{\dagger}}$ is the adjoint Fokker-Planck operator,
$\text{\ensuremath{\boldsymbol{\epsilon}} is a diagonal matrix of the parities }\epsilon_{i}=\pm1$
of $x_{i}$ under time reversal, and the stationary distribution is
time-reversal invariant, $P(\boldsymbol{x})=P(\boldsymbol{\epsilon x})$.
This implies Onsager-Casimir symmetry $\boldsymbol{\epsilon}(\boldsymbol{\lambda c})=(\boldsymbol{\lambda c})^{\text{tr}}\boldsymbol{\epsilon}$
for the regression matrix and $\boldsymbol{\epsilon c}=\boldsymbol{c\epsilon}$
for the covariance matrix. The matrix of covariances is then determined
as
\[
\boldsymbol{c}=\left(\boldsymbol{\lambda}^{\mathrm{ir}}\right)^{-1}\left(\boldsymbol{\sigma}\boldsymbol{\sigma}^{\text{tr}}\right),
\]
where $\lambda_{ij}^{\mathrm{ir}}=(\lambda_{ij}+\epsilon_{i}\epsilon_{j}\lambda_{ij})$
.

The Gauss-Markov property of the Ornstein-Uhlenbeck process ensures
that the correlation function
\begin{equation}
\boldsymbol{C}(t-t^{\prime})\equiv\langle\boldsymbol{x}(t)\boldsymbol{x}^{\text{tr}}(t^{\prime})\rangle=e^{-\boldsymbol{\lambda}|\Delta t|}\boldsymbol{c},\mathbf{}\label{eq:autocorr}
\end{equation}
decays exponentially and that its Fourier transform, the power spectral
density
\begin{equation}
\boldsymbol{C}(\Omega)=(\text{\textminus}i\Omega\boldsymbol{1}+\boldsymbol{\lambda})^{-1}\left(2\boldsymbol{D}\right)(i\Omega\boldsymbol{1}+\boldsymbol{\lambda^{\text{tr}}})^{-1},\label{eq:spectral-density}
\end{equation}
is a multivariate Lorentzian in the angular frequency $\Omega$ \cite{van1992stochastic}. 

In what follows, we shall take $\boldsymbol{\Lambda}$, the matrix
exponential of the mean regressions rates and $\boldsymbol{c}$, the
covariance matrix, to be the independent parameters. Estimates of
the parameters in the diffusion matrix can then be obtained from the
estimates of $\boldsymbol{\Lambda}$ and $\boldsymbol{c}$ through
the stationarity condition. Thus, there are $M^{2}+M$ independent
parameters for a $M$-variate Ornstein-Uhlenbeck process. For notational
brevity, the set of all unknown parameters is collected in $\boldsymbol{\theta}=(\boldsymbol{\Lambda},\boldsymbol{c})$. 

\section{Bayesian inference\label{sec:Bayesian-inference}}

\emph{Parameter Estimation: }Consider now the discrete time series
$\boldsymbol{X}=\big\{\boldsymbol{x}_{1},\boldsymbol{x}_{2},...,\boldsymbol{x}_{N}\big\}$,
consisting of $N$ observations of the sample path $\boldsymbol{x}(t)$
at the discrete times $t=n\Delta t$ with $n=1,\ldots,N.$ Each observation
$\boldsymbol{x}_{n}$ is an $M$-dimensional vector corresponding
to the number of components of the multivariate Ornstein-Uhlenbeck
process. From the Markov property of the process, the probability
of the path, given the parameters in $\boldsymbol{\theta}$, is
\begin{equation}
P\left(\boldsymbol{X}\vert\boldsymbol{\theta}\right)=\prod_{n=1}^{N-1}P_{1\vert1}\left(\boldsymbol{x}_{n+1}\vert\boldsymbol{x}_{n},\boldsymbol{\theta}\right)P_{1}\left(\boldsymbol{x}_{1}\vert\boldsymbol{\theta}\right).
\end{equation}
The probability $P\left(\boldsymbol{\theta}\vert\boldsymbol{X}\right)$
of the parameters, given the sample path, is given by Bayes theorem
to be
\begin{equation}
P\left(\boldsymbol{\theta}\vert\boldsymbol{X}\right)=\dfrac{P\left(\bm{X}\vert\boldsymbol{\theta}\right)P\left(\boldsymbol{\theta}\right)}{P\left(\boldsymbol{X}\right)}.\label{eq:bayes-theorem}
\end{equation}
The denominator $P\left(\bm{X}\right)$ is a normalization independent
of the parameters, and thus it can be ignored in parameter estimation.
Using informative uniform priors for $P\left(\boldsymbol{\theta}\right)$,
the logarithm of the posterior probability, after using the explicit
forms of $P_{1\vert1}$ and $P_{1}$, is in matrix form\begin{widetext}
\begin{align}
\ln & P\left(\bm{\theta}\vert\boldsymbol{X}\right)=-\tfrac{1}{2}(\boldsymbol{x}_{1}^{\text{tr}},\cdots,\boldsymbol{x}_{N}^{\text{tr}})\left(\begin{array}{cccccc}
\boldsymbol{a}'_{o} & \boldsymbol{a}_{1}\\
\boldsymbol{a}_{1} & \boldsymbol{a}{}_{o} & \boldsymbol{a}_{1}\\
 & \ddots & \ddots & \ddots\\
 &  & \ddots & \ddots & \ddots\\
 &  &  & \boldsymbol{a}_{1} & \boldsymbol{a}{}_{o} & \boldsymbol{a}_{1}\\
 &  &  &  & \boldsymbol{a}_{1} & \boldsymbol{a}''_{o}
\end{array}\right)\left(\begin{array}{c}
\boldsymbol{x}_{1}\\
\\
\vdots\\
\vdots\\
\\
\boldsymbol{x}_{N}
\end{array}\right)-\tfrac{N-1}{2}\ln|2\pi\boldsymbol{\Sigma}|-\tfrac{1}{2}\ln|2\pi\boldsymbol{c}|.\label{eq:posteriorBayesI}
\end{align}
where
\[
\boldsymbol{a}_{0}=\boldsymbol{\Sigma}^{-1}+\boldsymbol{\Lambda}{}^{\text{tr}}\boldsymbol{\Sigma}^{-1}\boldsymbol{\Lambda},\qquad\boldsymbol{a}_{1}=-\boldsymbol{\Sigma}^{-1}\boldsymbol{\Lambda},\qquad\boldsymbol{a}'_{0}=\boldsymbol{\Sigma}^{-1}+\boldsymbol{c}^{-1},\quad\boldsymbol{a}''_{0}=\boldsymbol{\Sigma}^{-1}.
\]
We note that the above matrix of covariances is real, symmetric, tridiagonal,
and almost-Toeplitz (only the first and last elements of the diagonal
differ from the remaining elements). Symmetry follows from the Gaussian
property \cite{rasmussen2006gaussian}, tridiagonality from the Markov
property \cite{kavcic2000matrices}, and the Toeplitz character from
stationarity \cite{van1992stochastic}. Exploiting these properties,
with some elementary manipulations detailed in Appendix \ref{app:MAP-estimates},
the posterior probability can be written in terms of the four matrix
sufficient statistics as
\begin{equation}
\ln P\left(\bm{\theta}\vert\bm{X}\right)=-\frac{\bm{\Sigma}^{-1}}{2}:\Big[\left(\bm{\Lambda}-\bm{T}_{2}\bm{T}_{3}^{-1}\right)\bm{T}_{3}\left(\bm{\Lambda}-\bm{T}_{2}\bm{T}_{3}^{-1}\right)^{\text{tr}}+\left(\bm{T}_{1}-\bm{T}_{2}\bm{T}_{3}^{-1}\bm{T}_{2}^{\text{tr}}\right)\Big]-\tfrac{N-1}{2}\ln|2\pi\boldsymbol{\Sigma}|-\tfrac{\bm{c}^{-1}}{2}:\boldsymbol{T}_{4}-\tfrac{1}{2}\ln|2\pi\boldsymbol{c}|.\label{eq:postBayesI}
\end{equation}
\end{widetext}The sufficient statistics matrices are\begin{subequations}\label{eq:sufficientStatistics}
\begin{flalign}
\boldsymbol{T}_{1} & =\sum_{n=1}^{N-1}\boldsymbol{x}_{n+1}\boldsymbol{x}_{n+1}^{\text{tr}},\quad\boldsymbol{T}_{2}=\sum_{n=1}^{N-1}\boldsymbol{x}_{n+1}\boldsymbol{x}_{n}^{\text{tr}},\\
\boldsymbol{T}_{3} & =\sum_{n=1}^{N-1}\boldsymbol{x}_{n}\boldsymbol{x}_{n}^{\text{tr}},\quad\quad\quad\boldsymbol{T}_{4}=\boldsymbol{x}_{1}\boldsymbol{x}_{1}^{\text{tr}}.
\end{flalign}
\end{subequations}We note that these are matrices of covariances
at equal times and one observation time apart. Therefore, they can
be evaluated at a cost proportional to $2N-1$. It is this property
that is results in our method being ``fast'', that is, requiring
only $O(N)$ computational cost. 

The MAP estimates for mean regression rates and the covariances turn
out to be\begin{subequations}\label{eq:mapBayesI}
\begin{align}
\bm{\bm{\Lambda}^{\ast}} & =\bm{T}_{2}\boldsymbol{T}_{3}^{-1},\label{eq:map_lambda}\\
\bm{\Sigma^{\ast}} & =\dfrac{1}{N}\left(\bm{T}_{1}-\boldsymbol{T}_{2}\boldsymbol{T}_{3}^{-1}\boldsymbol{T}_{2}^{\text{tr}}\right).\label{eq:map_sigma}
\end{align}
\end{subequations}The standard errors are obtained from the Hessian
matrix $\bm{A}$ of the posterior probability evaluated at the maximum
\cite{jeffreys1998theory,jaynes2003probability,sivia2006data}. Its
explicit form is provided in the Appendix \ref{app:Standard-errors-and}.
The four $M\times M$ sufficient statistic matrices $\boldsymbol{T}_{i}$
rather than the considerably larger $M\times N$ times series $\boldsymbol{X}$
contains all the information relevant for inference. Their use reduces
both the storage and computational cost of the inference algorithm. 

The estimate for the mean regression rate is recognized to be the
multivariate generalization of our earlier result for the univariate
Ornstein-Uhlenbeck process \cite{bera2017fast}. An explicit estimate
for $\boldsymbol{c}$ can be obtained from that of $\boldsymbol{\Sigma}$
when the Fokker-Planck operator obeys detailed balance. Then Onsager-Casimir
symmetry $\boldsymbol{\epsilon}(\boldsymbol{\lambda c})=(\boldsymbol{\lambda c})^{\text{tr}}\boldsymbol{\epsilon}$
implies $\boldsymbol{\Lambda}\boldsymbol{c}=\boldsymbol{c}\boldsymbol{\epsilon}\boldsymbol{\Lambda}^{\text{tr}}\boldsymbol{\epsilon}$
and, from the definition of $\boldsymbol{\Sigma}$, it follows that
the MAP estimate for $\boldsymbol{c}$ is
\begin{equation}
\boldsymbol{c}^{\ast}=\boldsymbol{\Sigma}^{\ast}\left[\boldsymbol{1}-(\boldsymbol{\Lambda}^{\ast}\boldsymbol{\epsilon})^{\text{tr}}(\boldsymbol{\Lambda}^{\ast}\boldsymbol{\epsilon})^{\text{tr}}\right]^{-1}.\label{eq:covariance-OC}
\end{equation}
The MAP estimate for the matrix of diffusion coefficients then follows
from the stationarity condition. This is the multivariate generalization
of our earlier result for the diffusion coefficient for the univariate
Ornstein-Uhlenbeck process. We refer to this method as ``Bayes I'',
following the terminology of \cite{bera2017fast}.

In the absence of detailed balance, such explicit expressions can
no longer be found and linear systems have to be solved to obtain
the MAP estimate of $\boldsymbol{c}$ from that of $\boldsymbol{\Sigma}$
and, then, to relate it to the matrix of diffusion coefficients. We
shall pursue this elsewhere. 

An alternative Bayesian procedure for directly estimating $\boldsymbol{c}$
is arrived at by interpreting the time series $\boldsymbol{X}$ as
independent repeated samples from the stationary distribution $\mathcal{N}\left(\boldsymbol{0},\boldsymbol{c}\right)$.
Using non-informative priors, the expression of the logarithm of the
posterior probability in this approach is given as
\begin{align}
\ln P\left(\boldsymbol{c}\,|\boldsymbol{X}\right) & =\dfrac{N}{2}\ln\dfrac{1}{\left(2\pi\right)^{M}|\bm{c}|}-\dfrac{1}{2}\sum_{n=1}^{N}\boldsymbol{x}_{n}^{\text{tr}}\boldsymbol{c}^{-1}\,\boldsymbol{x}_{n},\label{eq:posteriorBayesII}
\end{align}
from which the MAP estimate
\begin{align}
\boldsymbol{c}^{\ast}=\frac{1}{N}\sum_{n=1}^{N}\boldsymbol{x}_{n}\boldsymbol{x}_{n}^{\text{tr}}=\frac{1}{N}\boldsymbol{T}_{3},\label{eq:mapBayesII}
\end{align}
follows straightforwardly. The Bayesian inference method described
above, using the stationary distribution of the Ornstein-Uhlenbeck
process, is referred to as \textquotedblleft Bayes II\textquotedblright . 

Bayes I exploits both the Gaussian and Markovian character of the
process, while Bayes II exploits only its Gaussian character. Therefore,
numerical agreement between the above two methods of estimating the
covariance matrix provides a stringent test of the stationary, Gaussian,
and Markovian characters of the process. Since the Ornstein-Uhlenbeck
process is the only continuous process with all of these properties,
agreement affirms it as the data generating model. The preceding equations
(\ref{eq:posteriorBayesI}-\ref{eq:mapBayesII}) are the main results
of this paper.

\emph{Model comparison: }Thus far we assumed that the data generating
model was given and that only the parameters of the model needed to
be estimated. In certain circumstances, though, the model itself may
be uncertain and it becomes necessary to estimate the probability
of different models $\mathcal{M}_{\alpha}$ \cite{jeffreys1998theory,kashyap1977bayesian,mackay1992bayesian,zellner1984,gregory2005bayesian}.
The probability of a model, given the data, is
\begin{equation}
P(\mathcal{M}_{\alpha}|\boldsymbol{X})\propto P(\boldsymbol{X}|\mathcal{M}_{\alpha})P(\mathcal{M}_{\alpha}),
\end{equation}
where the first term on the right is the ``evidence'' of the model
and the second term is the prior probability of the model. We shall
assume all models to be, a priori, equally likely. The evidence is
the normalizing constant in Eq.(\ref{eq:bayes-theorem}), given as
an integral over the space of parameters $\boldsymbol{\theta}$ contained
in the drift and diffusion matrices:
\begin{equation}
P(\boldsymbol{X}|\mathcal{M}_{\alpha})=\int P(\boldsymbol{X}|\boldsymbol{\theta},\mathcal{M}_{\alpha})P(\boldsymbol{\theta}|\mathcal{M}_{\alpha})d\boldsymbol{\theta}.
\end{equation}
For unimodal posterior distributions, the height at the MAP value
$\boldsymbol{\theta}^{\ast}$ times the width $\Delta\boldsymbol{\theta}$
of the distribution is, often, a very good approximation for the evidence,
\begin{equation}
P(\boldsymbol{X}|\mathcal{M}_{\alpha})\simeq P(\boldsymbol{X}|\boldsymbol{\theta}^{\ast},\mathcal{M}_{\alpha})P(\boldsymbol{\theta}^{\ast}|\mathcal{M}_{\alpha})\Delta\boldsymbol{\theta}.\label{eq:evidence}
\end{equation}
The first term is the best fit likelihood while the second term, the
product of the prior for the MAP estimate and the standard error of
this estimate is called the Ockham factor. Thus models that achieve
a compromise between the degree of fit to the data and the number
of parameters required for the fit are ones that are favoured by the
Bayesian model selection procedure. This avoids the over-fitting that
would occur if the degree of fit was made the sole criterion for model
selection, and it encodes the commonsense ``principle of parsimony'',
 attributed to William of Ockham, which states that between two models
that fit the data equally well, the simpler one is to be preferred.
Both the evidence and the Ockham factor can be obtained straightforwardly
for the Ornstein-Uhlenbeck models, and we shall make use of this below
for model selection within a family of Ornstein-Uhlenbeck models. 

To summarize, we derive two $O(N)$ methods to compute the parameters
of a multivariate Ornstein-Uhlenbeck process in terms of sufficient
statistics. The cost of computing these statistics for an additional
$N^{\prime}$ data points is only $O(N^{\prime})$. Thus previous
computations can be reused with the arrival of fresh data, making
the method suitable for ``online'' estimation. The fact that new
data can be incrementally added to the sufficient statistics is also
useful for comparisons within a family of $M$-variate Ornstein-Uhlenbeck
models. For large $M$, the determinant of the Hessian matrix needed
for model comparison can be computed efficiently using iterative Krylov
subspace methods, thereby allowing for a full ``online'' inference
of both models and their parameters. 

\section{Path sampling\label{sec:Path-sampling}}

The solution of the Fokker-Planck equation, Eq.(\ref{eq:conditional-probability}),
provides a method for sampling paths of the multivariate Ornstein-Uhlenbeck
process \emph{exactly}. Given an initial state $\boldsymbol{x}$ at
time $t$ and final state $\boldsymbol{x}^{\prime}$ at time $t^{\prime}$
, the quantity $\boldsymbol{x}^{\prime}-\boldsymbol{\Lambda}\boldsymbol{x}$
is normally distributed with mean zero and variance $\boldsymbol{\Sigma}$,
a property that was first recognized by Uhlenbeck and Ornstein. Therefore,
a sequence of states at times $t=n\Delta t$, where $n$ is a positive
integer, forming a discrete sampling of a path can be obtained from
the following iteration \cite{uhlenbeck1930theory}
\begin{equation}
\boldsymbol{x}_{n+1}=\boldsymbol{\Lambda}\boldsymbol{x}_{n}+\sqrt{\boldsymbol{\Sigma}}\,\boldsymbol{\xi}_{n},\label{eq:path-sampling}
\end{equation}
where $\sqrt{\boldsymbol{\Sigma}}$ is a matrix square-root of $\boldsymbol{\Sigma}$,
and $\boldsymbol{\xi}_{n}\sim\mathcal{N}(\boldsymbol{0},\boldsymbol{1})$
is an $M$-dimensional uncorrelated normal variate with zero mean
and unit variance. The exponential of the regression matrix and the
square-root of the variance matrix are the two key quantities in the
iteration. They can be obtained analytically in low dimensional problems,
but for high-dimensional problems they will, in general, have to be
obtained numerically \cite{parker2012sampling}. The sampling interval
must satisfy $\lambda_{\mathrm{max}}\Delta t\ll1$ such that the shortest
time scale in the dynamics, corresponding to the inverse of the largest
eigenvalue $\lambda_{\mathrm{max}}$ of the regression matrix, is
resolved in the samples. In the following section, we use the above
method to sample paths of the Brownian harmonic oscillator, which
is equivalent to a bivariate Ornstein-Uhlenbeck process, to test the
accuracy of our Bayesian estimates.

\section{Brownian harmonic oscillator\label{sec:Brownian-harmonic-oscillator}}

\emph{Inference problem: }We now apply the results above to the physically
important case of a massive Brownian particle confined in a harmonic
potential described by the Langevin equation
\begin{equation}
m\dot{v}+\gamma v+\nabla U(x)=\xi.\label{eq:langevin}
\end{equation}
Here the pair $(x,v)$ describes the state of the particle in its
phase space of position and velocity, while $m$ and $\gamma$ are
the particle mass and friction coefficient respectively. The potential
$U=\frac{1}{2}kx^{2}$ is harmonic with a stiffness $k$. $\xi\left(t\right)$is
a zero-mean Gaussian white noise with variance $\langle\xi\left(t\right)\xi\left(t'\right)\rangle=2k_{B}T\gamma\delta\left(t-t'\right)$
that satisfies the fluctuation-dissipation relation \cite{chandrasekhar1943stochastic}.
The inference problem is to jointly estimate the triplet of parameters
$(m,\gamma,k)$ from discrete observations of the position and velocity
and to estimate the correlation functions and the spectral densities
from these observations. 

\emph{Bivariate Ornstein-Uhlenbeck process: }The Langevin equation
can be recast as a bivariate Ornstein-Uhlenbeck process in phase space
for the pair ($x,v$) as
\begin{equation}
dv=\left(-\omega_{0}^{2}x-v/\tau\right)\,dt+\sigma dW_{v},\qquad dx=v\,dt.
\end{equation}
Here $\omega_{0}^{2}=k/m$ is the natural frequency of the undamped
harmonic oscillator, $\tau=m/\gamma$ is the characteristic time scale
associated with the thermalization of the momentum due to viscous
dissipation, and $\sigma=\sqrt{2D}$, where the diffusion coefficient
$D$ of the particle is defined, as usual, by the Einstein relation
$D=k_{B}T\gamma^{-1}$ \cite{einstein1905theory,kubo1966fluctuation}.
Thus, we have constructed three independent parameters from the four
dependent parameters. The resulting bivariate system can be written
in matrix form as
\begin{equation}
d\left(\begin{array}{c}
x\\
v
\end{array}\right)=-\bm{\lambda}\left(\begin{array}{c}
x\\
v
\end{array}\right)dt+\boldsymbol{\sigma}\left(\begin{array}{c}
dW_{x}\\
dW_{v}
\end{array}\right),\label{eq:phase_space}
\end{equation}
where the mean regression matrix is
\begin{equation}
\bm{\lambda}=\begin{pmatrix}0 & -1\\
\omega_{0}^{2} & 1/\tau
\end{pmatrix},
\end{equation}
and the volatility matrix is
\begin{equation}
\bm{\sigma=}\begin{pmatrix}0 & 0\\
0 & \sqrt{2D}/\tau
\end{pmatrix}.
\end{equation}
The structure of the volatility matrix ensures that the positional
Wiener process $dW_{x}$ does not enter the dynamics. 

At thermal equilibrium, the joint distribution of position and velocity
factorize into the Gibbs distribution for the position and the Maxwell-Boltzmann
distribution for the velocity to give a diagonal covariance matrix
\begin{equation}
\bm{c}=\begin{pmatrix}k_{B}T/k & 0\\
0 & k_{B}T/m
\end{pmatrix}.\label{eq:covarianceDiagonal}
\end{equation}
It is easily verified that the stationarity condition, Eq. (\ref{eq:stationarity}),
is satisfied by the above matrices. The condition of micro-reversibility
translates, here, into Onsager-Casimir symmetry, $\boldsymbol{\lambda}^{\mathrm{ir}}\boldsymbol{c}=\boldsymbol{\sigma}\boldsymbol{\sigma}^{\text{tr}}$,
where $\lambda_{ij}^{\mathrm{ir}}=(\lambda_{ij}+\epsilon_{i}\epsilon_{j}\lambda_{ij})$
is the irreversible part of the drift coefficient for variables that
are, respectively, even or odd under time reversal. Then, the only
non-zero entry of $\boldsymbol{\lambda}^{\mathrm{ir}}$ is $\lambda_{22}^{\mathrm{ir}}=\tau^{-1}$,
and it is trivial to verify the Onsager-Casimir symmetry.

\emph{Path sampling: }To sample paths of the Brownian harmonic oscillator
exactly, it is necessary to obtain the exponential of the regression
matrix $\boldsymbol{\lambda}$ and the square-root of the variance
matrix $\boldsymbol{\Sigma}$. From the Cayley-Hamilton theorem, the
former is easily found to be
\begin{equation}
\boldsymbol{\Lambda=}e^{-\boldsymbol{\lambda}\Delta t}=\Lambda_{1}\boldsymbol{1}+\Lambda_{2}\boldsymbol{\lambda},\label{eq:cayHam}
\end{equation}
while the latter is obtained from a Cholesky factorization of $\boldsymbol{\Sigma}$
into a lower triangular matrix and its transpose,
\begin{equation}
\sqrt{\boldsymbol{\Sigma}}=\begin{pmatrix}s_{1} & 0\\
s_{2} & s_{3}
\end{pmatrix}.\label{eq:choleskyF}
\end{equation}
See appendix \ref{app:pathSampling} for the derivation of the coefficients.
\begin{table*}
\begin{tabular}{|>{\centering}p{2cm}|>{\centering}p{2cm}|>{\centering}p{2cm}|}
\hline 
\multicolumn{3}{|c|}{$m\;\big(ng\big)$}\tabularnewline
\hline 
Simulation & Bayes I & Bayes II\tabularnewline
\hline 
\hline 
$1$ & $0.994\pm0.006$ & $0.993\pm0.008$\tabularnewline
\hline 
$1.5$ & $1.521\pm0.014$ & $1.517\pm0.012$\tabularnewline
\hline 
$2$ & $2.107\pm0.015$ & $2.092\pm0.016$\tabularnewline
\hline 
\end{tabular}%
\begin{tabular}{>{\centering}p{2cm}|>{\centering}p{2cm}|>{\centering}p{2cm}|>{\centering}p{2cm}|>{\centering}p{2cm}|}
\hline 
\multicolumn{3}{c|}{$k\;(mg/s^{2})$} & \multicolumn{2}{c|}{$\gamma\;(\mu g/s^{2})$}\tabularnewline
\hline 
Simulation & Bayes I & Bayes II & Simulation & Bayes I\tabularnewline
\hline 
\hline 
$225$ & $224.81\pm1.72$ & $224.49\pm1.75$ & $3$ & $2.966\pm0.021$\tabularnewline
\hline 
$250$ & $251.97\pm1.93$ & $251.82\pm1.97$ & $1.936$ & $1.974\pm0.016$\tabularnewline
\hline 
$300$ & $296.51\pm2.37$ & $298.44\pm2.46$ & $1.639$ & $1.589\pm0.015$\tabularnewline
\hline 
\end{tabular}

\caption{Bayesian MAP estimates and standard errors of the parameters of the
Brownian harmonic oscillator. There is excellent agreement between
the two Bayesian methods and with the parameter values used to generate
the paths. \label{tab:springconstant}}
\end{table*}

\emph{Parameter estimation: }We now take the time series $\boldsymbol{X}=\big\{(x_{1},v_{1})^{\text{tr}},\ldots,(x_{N},v_{N})^{\text{tr}}\big\}$
of discrete observations of positions and velocities obtained from
the exact path sampling and construct from it the sufficient statistics\begin{subequations}\label{eq:bho-T-matrices}
\begin{alignat}{1}
\boldsymbol{T}_{1} & =\sum_{n=1}^{N-1}\begin{pmatrix}x_{n+1}^{2} & x_{n+1}v_{n+1}\\
v_{n+1}x_{n+1} & v_{n+1}^{2}
\end{pmatrix},\\
\boldsymbol{T}_{2} & =\sum_{n=1}^{N-1}\begin{pmatrix}x_{n+1}x_{n} & x_{n+1}v_{n}\\
v_{n+1}x_{n} & v_{n+1}v_{n}
\end{pmatrix},\\
\boldsymbol{T}_{3} & =\sum_{n=1}^{N-1}\begin{pmatrix}x_{n}^{2} & x_{n}v_{n}\\
v_{n}x_{n} & v_{n}^{2}
\end{pmatrix}.
\end{alignat}
\end{subequations}From these, we compute the MAP estimate for the
regression matrix,
\[
\boldsymbol{\lambda}^{\ast}=-\frac{1}{\Delta t}\ln\boldsymbol{\Lambda}^{\ast}=-\frac{1}{\Delta t}\ln(\bm{T}_{2}\boldsymbol{T}_{3}^{-1}),
\]
and the covariance matrix,
\[
\boldsymbol{c}^{\ast}=\dfrac{1}{N}\left(\bm{T}_{1}-\boldsymbol{T}_{2}\boldsymbol{T}_{3}^{-1}\boldsymbol{T}_{2}^{\text{tr}}\right)\left[\boldsymbol{1}-(\bm{T}_{2}\boldsymbol{T}_{3}^{-1}\boldsymbol{\epsilon})^{\text{tr}}(\bm{T}_{2}\boldsymbol{T}_{3}^{-1}\boldsymbol{\epsilon})^{\text{tr}}\right]^{-1}
\]
Using the above, $\boldsymbol{\lambda}^{*}$ yields the MAP estimates
for the natural frequency and the relaxation time scale
\begin{equation}
\left(\frac{k}{m}\right)^{\ast}=\lambda_{21}^{\ast},\quad\left(\frac{\gamma}{m}\right)^{\ast}=\lambda_{22}^{\ast},
\end{equation}
while $\boldsymbol{c}^{*}$ yields the MAP estimate of the spring
constant and the mass, in units of $k_{B}T$
\begin{equation}
\frac{k_{B}T}{k^{*}}=c_{11}^{\ast},\qquad\frac{k_{B}T}{m^{*}}=c_{22}^{\ast}.
\end{equation}
The friction constant is estimated by eliminating the mass between
two of the previous ratios,
\begin{equation}
\gamma^{\ast}=\frac{k_{B}T\lambda_{22}^{\ast}}{c_{22}^{\ast}}.
\end{equation}
The three preceding equations provide the Bayes I map estimates of
the oscillator parameters.
\begin{figure}
\includegraphics[width=0.46\textwidth]{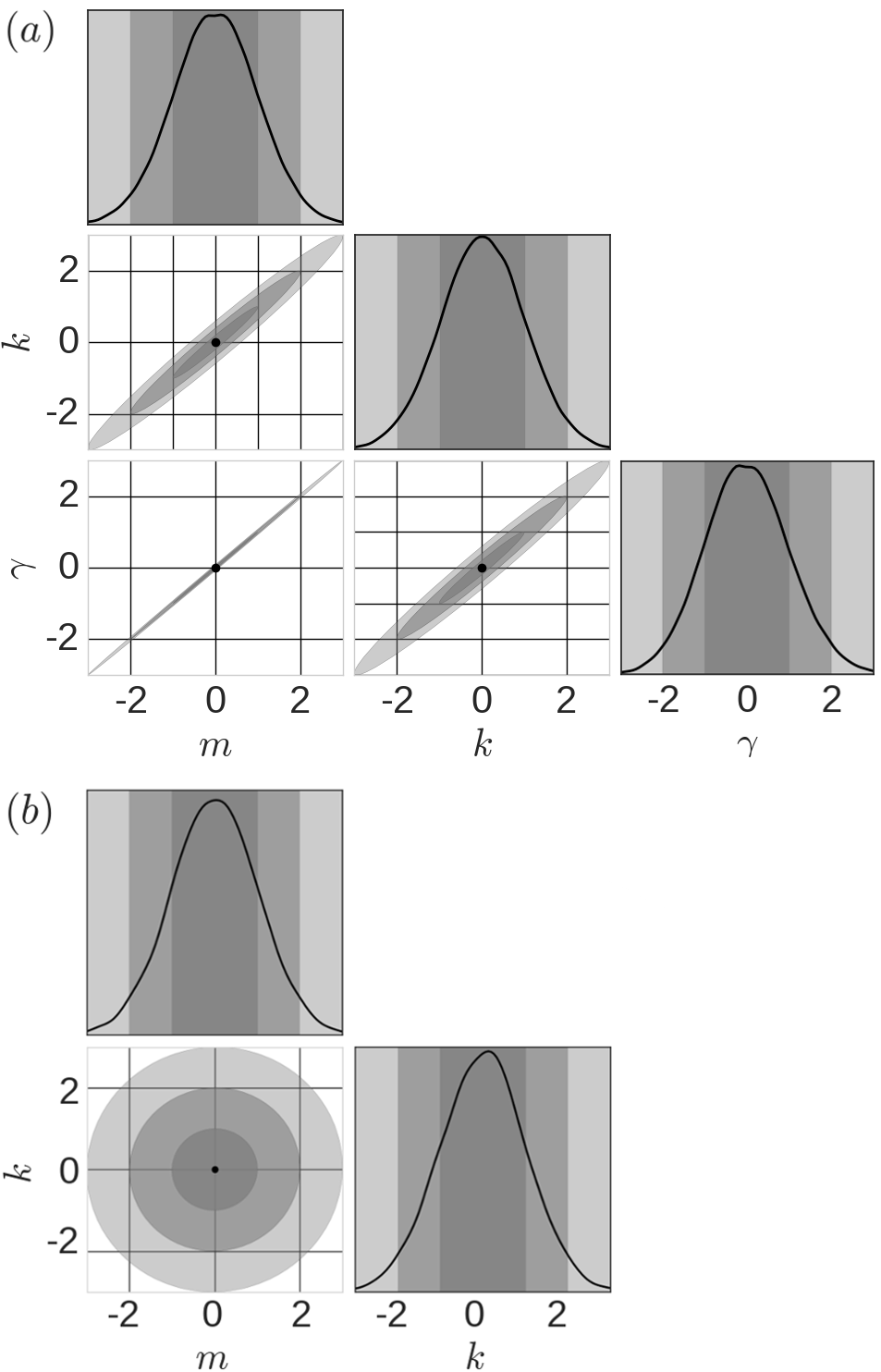}\caption{Corner plots of the joint posterior distribution of the mass $m$,
friction $\gamma$ and stiffness $k$. Panel $(a)$ shows Bayes I,
Eq.(\ref{eq:mapBayesI}) while panel $(b)$ shows Bayes II, Eq.(\ref{eq:map_stationary}).
The variables have been scaled as $\left(x_{i}-\mu_{_{i}}\right)/\sigma_{_{ii}}$,
where $\mu_{i}$ and $\sigma_{ii}^{2}$ are the MAP estimate and variance
of $x_{i}$ respectively. The MAP estimate is marked by a black dot,
and the regions of $70\%$, $95\%$ and $99\%$ posterior probability
have been shaded. \label{fig:inference}\label{fig:stationary}}
 
\end{figure}

Next we use Eq.(\ref{eq:covarianceDiagonal}) for the covariance matrix
in the bivariate analog of Eq.(\ref{eq:posteriorBayesII}) for the
logarithm of the posterior probability. The MAP estimates and the
error bars for mass $m$ and spring constant $k$ are then\begin{subequations}\label{eq:map_stationary}
\begin{align}
\dfrac{k^{\ast}}{k_{B}T}=\dfrac{N}{\sum_{n=1}^{N}x_{n}^{2}} & ,\qquad\sigma_{k}=\frac{\sqrt{2}}{\sqrt{N}}k^{\ast},\label{eq:map_k}\\
\dfrac{m^{\ast}}{k_{B}T}=\dfrac{N}{\sum_{n=1}^{N}v_{n}^{2}}, & \qquad\,\,\sigma_{m}=\frac{\sqrt{2}}{\sqrt{N}}m^{\ast}.\label{eq:map_v}
\end{align}
\end{subequations}These correspond to Bayes II estimates of the oscillator
parameters. 
\begin{figure}
\includegraphics[width=0.485\textwidth]{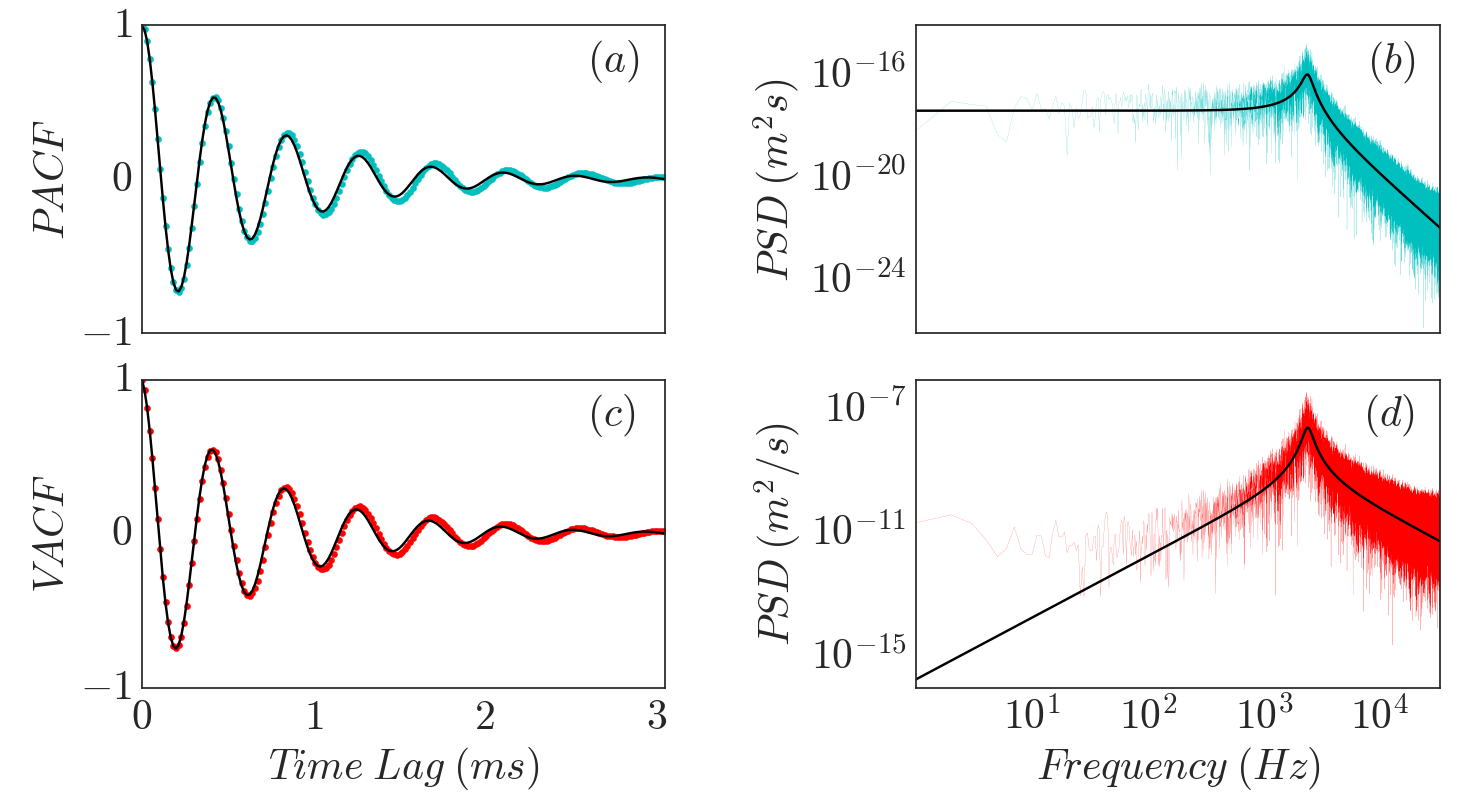}

\caption{Bayesian estimates of the autocorrelation function and spectral density
of position are shown in (a) and (b) respectively as solid lines.
There is excellent agreement with simulations shown in cyan. Corresponding
estimates for the velocity are shown in (c) and (d) with simulations
shown in red. \label{fig:autocorrelation} }
\end{figure}

In Table \ref{tab:springconstant}, we provide the MAP estimates and
corresponding error bars for three different time-series data obtained
from the simulation of the underdamped Brownian harmonic oscillator.
This clearly shows that the MAP estimates of both Bayesian methods
and values of the simulation parameters are in excellent agreement.
The error bars have been calculated from the Hessian matrix. The Bayesian
standard error in estimating the parameters is less than $1\%$ for
each case.

In Fig.(\ref{fig:inference}) we show the posterior distribution of
the parameters as corner plots \cite{foreman2016corner}, with panel
(a) corresponding to Bayes I and panel (b) to Bayes II. The distributions
have been shifted to the MAP estimates, which, therefore are always
at the origin marked by the black dot, and scaled by the variances.
The Bayesian credible intervals corresponding to $70\%$, $95\%$,
and $99\%$ of the posterior probability are shown in shades of gray. 

\emph{Correlation functions and spectral densities: }The MAP estimates
of the parameters provide a novel way of estimating the correlation
functions and spectral densities. Their expressions can be obtained
from Eq.(\ref{eq:autocorr}) and Eq.(\ref{eq:spectral-density}) using
the explicit forms of the covariance and the regression matrices.
The autocorrelations are
\begin{align*}
\langle x\left(t\right)x\left(0\right)\rangle= & \dfrac{k_{B}T}{k}\exp\left(\dfrac{-t}{2\tau}\right)\dfrac{(2\omega\tau\cos\omega t+\sin\omega t)}{2\omega\tau},\\
\langle v\left(t\right)v\left(0\right)\rangle= & \dfrac{k_{B}T}{m}\exp\left(\dfrac{-t}{2\tau}\right)\dfrac{(2\omega\tau\cos\omega t-\sin\omega t)}{2\omega\tau}.
\end{align*}
and the spectral densities are\begin{subequations}
\begin{align}
C_{xx}(\Omega)= & \dfrac{2\gamma k_{B}T}{m^{2}(\omega_{0}^{2}-\Omega^{2})^{2}+\gamma^{2}\Omega^{2}},\\
C_{vv}(\Omega)= & \dfrac{2\gamma k_{B}T\Omega^{2}}{m^{2}(\omega_{0}^{2}-\Omega^{2})^{2}+\gamma^{2}\Omega^{2}}.
\end{align}
\end{subequations}These expressions are evaluated at the Bayesian
MAP estimates for the parameters. In Fig.(\ref{fig:autocorrelation}),
the result is compared with the autocorrelation function and the discrete
Fourier transform of the time series. There is excellent agreement
between the two autocorrelations. We emphasize that no numerical function
fitting is required to obtain this estimate. The spectral density
is, arguably, even more impressive as it interpolates through the
noisy discrete Fourier transform in a sensible manner. Our estimation
of the spectral density shares a methodological similarity with Burg's
classic maximum-entropy method \cite{burg1967maximum}. The principal
differences are (i) that we choose the Ornstein-Uhlenbeck process
as the data generating model, rather than the discrete autoregressive
process assumed by Burg and (ii) that we work directly on the space
of trajectories rather than in the space of correlations. From a Bayesian
perspective, our estimation procedure encodes the prior information
that the data are generated by an Ornstein-Uhlenbeck process and,
therefore, will outperform all methods \cite{berg2004power,tassieri2012microrheology}
that do not incorporate this prior information, such as the one in
\cite{Bera:16}.
\begin{figure}[t]
\includegraphics[width=0.46\textwidth]{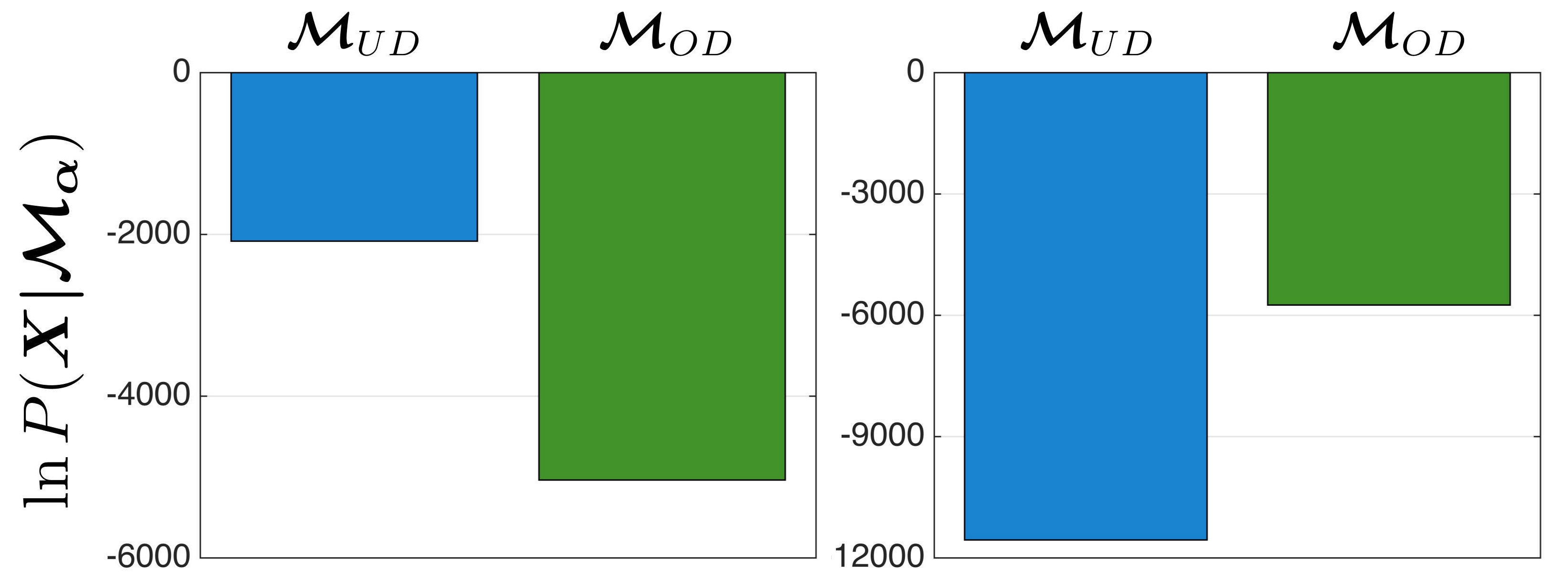}

\caption{Bayesian model selection procedure for time series data from underdamped
($\mathcal{M}_{UD}$) and overdamped ($\mathcal{M}_{OD}$) Brownian
harmonic oscillator. The logarithm of Bayesian evidence has been plotted
in the natural log units (nits). The plot for underdamped data $(\gamma=\gamma_{c}/10)$
is shown on the left, while the right panel is for overdamped data
$(\gamma=10\gamma_{c})$.\label{fig:modelSelection} }
\end{figure}

\emph{Model comparison: }To observe the inertial motion of a colloidal
particle in an optical trap, it is necessary to observe the motion
at time intervals much smaller than the momentum relaxation time,
that is, to ensure $\Delta t\ll\tau$. In the opposite limit, of $\Delta t\gg\tau$,
the inertia of the colloid is no longer relevant, and only purely
diffusive motion can be observed. These correspond, respectively,
to the Kramers and Smoluchowski limits of the Brownian harmonic oscillator.
In experiment, it is often not a priori clear if the observational
time scale places the system in the Kramers, Smoluchowski, or crossover
regime. A Bayesian model comparison provides a principled way to answer
this question, as we show below. 

We consider two Ornstein-Uhlenbeck data generating models
\begin{align}
\mathcal{M}_{UD}:\longrightarrow & \quad m\dot{v}+\gamma v+\nabla U=\xi\nonumber \\
\mathcal{M}_{OD}:\longrightarrow & \quad\gamma\dot{x}+\nabla U=\xi
\end{align}
corresponding, respectively, to motion on the Kramers\textbf{\emph{
}}(\textbf{u}nder\textbf{d}amped) and Smoluchowski (\textbf{o}ver\textbf{d}amped)
time scales. The latter is the formal $m\rightarrow0$ limit of Eq.(\ref{eq:langevin})
and can be obtained systematically by adiabatically eliminating the
velocity as a fast variable \cite{gardiner1984adiabatic}. The overdamped
oscillator is an univariate Ornstein-Uhlenbeck model with two parameters,
$\gamma$ and $k$, and Bayesian parameter estimation for it was presented
in \cite{bera2017fast}. The two-parameter overdamped model is simpler
than the three-parameter underdamped model. The question we try to
answer is which of these models provides the best explanation of the
data for the least number of parameters. 

To do so, we compare the posterior probability of the model, given
the data, by approximating the model evidence in terms of the maximum
likelihood and the Ockham factors, assuming equal prior probabilities
for the models as in section \ref{sec:Bayesian-inference}. The evidence
for both models is straightforwardly computed from the general expression
in the Appendix \ref{app:Standard-errors-and}. For the overdamped
model the sufficient statistics are the scalars
\begin{alignat}{1}
T_{1} & =\sum_{n=1}^{N-1}x_{n+1}^{2},\,\,T_{2}=\sum_{n=1}^{N-1}x_{n+1}x_{n},\,\,T_{3}=\sum_{n=1}^{N-1}x_{n}^{2}.
\end{alignat}
as was first pointed out in \cite{bera2017fast}. 

In Fig.(\ref{fig:modelSelection}), we plot the logarithm of the Bayesian
evidence $\ln P(\boldsymbol{X}|\mathcal{M}_{\alpha})$ in natural
log units (nits) for overdamped and underdamped Brownian motion. The
panel on the left contains the plot of the logarithm of the evidence
for underdamped data while the right panel is for overdamped data.
The comparison of $\ln P(\boldsymbol{X}|\mathcal{M}_{\alpha})$ for
the overdamped and underdamped models clearly shows that the evidence
is greater in each case for the true model of the data. 

\section{Summary\label{sec:conclusion}}

In summary, we have presented two Bayesian methods for inferring the
parameters of a multivariate Ornstein-Uhlenbeck process, given discrete
observations of the sample paths. An exact path sampling procedure
has been presented and utilized to validate the Bayesian methods for
the Brownian harmonic oscillator. The problem of Bayesian model comparison
has been addressed and applied to select between Kramers and Smoluchowski
limits of the Brownian harmonic oscillator. Future work will address
the problem of parameter inference and model selection when either
or both of the Markov and Gaussian properties of the process are relaxed.
\begin{acknowledgments}
We thank the two anonymous referees for their constructive criticism,
which led to an improvement in the presentation of our results. RA
thanks the Isaac Newton Trust for an Early Career Support grant. RS
is funded by a Royal Society-SERB Newton International Fellowship.
\end{acknowledgments}

\appendix
\widetext

\section{MAP estimates\label{app:MAP-estimates}}

In this appendix, we make explicit the derivation of Eq.(\ref{eq:postBayesI})
from Eq.(\ref{eq:posteriorBayesI}). To this end, we consider the
quadratic form $\boldsymbol{\Delta}_{n}^{\text{tr}}\bm{\Sigma^{-1}\Delta_{n}}=\left(\bm{x_{n+1}}-\bm{\Lambda}\bm{x_{n}}\right)^{\text{tr}}\bm{\Sigma^{-1}}\left(\bm{x_{n+1}}-\bm{\Lambda}\bm{x_{n}}\right)$
that occurs in the expression of the logarithm of the posterior probability
in Eq.(\ref{eq:posteriorBayesI}). Expanding terms and completing
the summation we have the following identity
\begin{align}
 & -\frac{1}{2}\sum_{n=1}^{N-1}\boldsymbol{\Delta}_{n}^{\text{tr}}\bm{\Sigma^{-1}\Delta_{n}}=-\dfrac{1}{2}\bm{\Sigma}^{-1}:\Big[\left(\bm{\Lambda}-\bm{T_{2}T_{3}^{-1}}\right)\left(\bm{T_{3}}\right)\left(\bm{\Lambda}-\bm{T_{2}T_{3}^{-1}}\right)^{\text{tr}}+\left(\bm{T}_{1}-\bm{T}_{2}\bm{T}_{3}^{-1}\bm{T}_{2}^{\text{tr}}\right)\Big].\label{eq:lp}
\end{align}
Here, we have used the definitions of sufficient statistics in Eq.(\ref{eq:sufficientStatistics}).
With this identification, the expression of the logarithm of the posterior
probability in terms of the sufficient statistics is given in Eq.(\ref{eq:postBayesI}).
Inspecting the expression, we recognize that the posterior is normal
in $\boldsymbol{\Lambda}$ which directly yields its MAP estimate
$\boldsymbol{\Lambda}^{*}=\bm{T}_{2}\bm{T}_{3}^{-1}$. Taking the
derivative of the above equation with respect to $\boldsymbol{\Sigma}^{-1}$
and using the matrix identity $\partial\ln\left(\det\bm{A}\right)/\partial\bm{A}=\left(\bm{A}^{\text{tr}}\right)^{-1}$,
we obtain
\begin{align}
 & \dfrac{\partial\ln P\left(\bm{\theta}\vert\bm{X}\right)}{\partial\bm{\Sigma^{-1}}}=-\dfrac{1}{2}\Big[\left(\bm{\Lambda}-\bm{T}_{2}\bm{T}_{3}^{-1}\right)\left(\bm{T}_{3}\right)\left(\bm{\Lambda}-\bm{T}_{2}\bm{T}_{3}^{-1}\right)^{\text{tr}}+\left(\bm{T}_{1}-\bm{T}_{2}\bm{T}_{3}^{-1}\bm{T}_{2}^{\text{tr}}\right)\Big]+\dfrac{N}{2}\bm{\Sigma}.
\end{align}
At the maximum, we obtain the MAP estimate as $\bm{\Sigma^{\ast}}=\dfrac{1}{N}\left(\bm{T}_{1}-\boldsymbol{T}_{2}\boldsymbol{T}_{3}^{-1}\boldsymbol{T}_{2}^{\text{tr}}\right).$
The estimates of the parameters of the underdamped oscillator follow
from $\bm{\Lambda^{\ast}}$ and $\bm{\Sigma^{\ast}}$ after algebraic
manipulations, see Appendix of \cite{rasmussen2006gaussian} for further
details. 

\section{Standard errors and evidence\label{app:Standard-errors-and}}

We now obtain an explicit expression for the standard errors and evidence
from the derivatives of the logarithm of the posterior distribution.
The second partial derivatives of the logarithm of the posterior distribution
with respect to $\boldsymbol{\Sigma}^{-1}$ and $\boldsymbol{\Lambda}$,
at the maximum, are\begin{subequations}
\begin{align}
\dfrac{\partial^{2}\ln P\left(\bm{\theta}\vert\bm{X}\right)}{\partial(\bm{\Sigma^{-1}})^{2}}= & -\dfrac{N}{2}\left(\bm{\Sigma}^{*}\right)^{2},\\
\dfrac{\partial^{2}\ln P\left(\bm{\theta}\vert\bm{X}\right)}{\partial\bm{\Lambda}^{2}}= & -\bm{\Sigma}^{*-1}(\bm{T_{3}}+N\boldsymbol{c}*)-\frac{N}{2}\bm{\Sigma}^{*-2}((\bm{\Lambda}^{*}\boldsymbol{c}^{*}+\boldsymbol{c}^{*}\bm{\Lambda}^{*\text{tr}}))^{2}.
\end{align}
\end{subequations}The mixed partial derivative at the maximum is
\begin{align}
\dfrac{\partial^{2}\ln P\left(\bm{\theta}\vert\bm{X}\right)}{\partial\bm{\Lambda}\partial\bm{\Sigma^{-1}}} & \eqsim-\frac{N}{2}(\bm{\Lambda}^{*}\boldsymbol{c}^{*}+\boldsymbol{c}^{*}\bm{\Lambda}^{*\text{tr}}).
\end{align}
The Hessian matrix, $\boldsymbol{A}=-\nabla\nabla\ln P\left(\bm{\theta}\vert\bm{X}\right)$,
at the maximum is then
\begin{align}
 & \boldsymbol{A}=\left(\begin{array}{ccc}
\tfrac{N}{2}(\bm{\Sigma}^{*})^{2} & \qquad & \frac{N}{2}(\bm{\Lambda}^{*}\boldsymbol{c}^{*}+\boldsymbol{c}^{*}\bm{\Lambda}^{*\text{tr}})\\
 & \qquad\\
\frac{N}{2}(\bm{\Lambda}^{*}\boldsymbol{c}^{*}+\boldsymbol{c}^{*}\bm{\Lambda}^{*\text{tr}})\, & \qquad & \bm{\Sigma}^{*-1}(\bm{T_{3}}-N\boldsymbol{c}^{*})-\frac{N}{2}\bm{\Sigma}^{*-2}(\bm{\Lambda}^{*}\boldsymbol{c}^{*}+\boldsymbol{c}^{*}\bm{\Lambda}^{*\text{tr}})^{2}
\end{array}\right).
\end{align}
The above expression of the Hessian matrix has been used to obtain
the standard errors of the MAP estimates and the evidence of a given
model. 

The expression of the Bayesian evidence of a model $\mathcal{M}_{\alpha}$
with several parameters is given in terms of the likelihood and the
Hessian matrix as \cite{mackay2003information}
\begin{equation}
P(\boldsymbol{X}|\mathcal{M}_{\alpha})\simeq P(\boldsymbol{X}|\boldsymbol{\theta}^{\ast},\mathcal{M}_{\alpha})\,P(\boldsymbol{\theta}^{\ast}|\mathcal{M}_{\alpha})[\det(\bm{A}/2\pi)]^{-1/2}.\label{eq:evidence-multivariate-1}
\end{equation}
This expression has been used to obtain the logarithm of the evidence
for a $M$-dimensional multivariate Ornstein-Uhlenbeck model $\mathcal{M}_{\alpha}$.
Using explicit forms of the likelihood and the Hessian matrix, at
the maximum, the expression of the logarithm of the evidence for a
model $\mathcal{M}_{\alpha}$ is given as
\begin{align}
\ln P\left(\bm{X}|\mathcal{M}_{\alpha}\right) & \simeq-\frac{N}{2}\left(\bm{\Sigma}^{*^{-1}}:\bm{\Sigma}^{*}+\ln((2\pi)^{M}\,|\bm{\Sigma}^{*}|)\right)-\frac{1}{2}\ln\Big(\det(\boldsymbol{A}_{22})\det(\boldsymbol{A}_{11}-\boldsymbol{A}_{12}\boldsymbol{A}_{22}^{-1}\boldsymbol{A}_{21})\Big)+\tfrac{M}{2}\ln2\pi.
\end{align}
Here $\boldsymbol{A}_{ij}$ are the elements of the Hessian matrix. 

\section{Path sampling\label{app:pathSampling}}

\begin{figure}
\textcolor{black}{\includegraphics[width=0.488\textwidth]{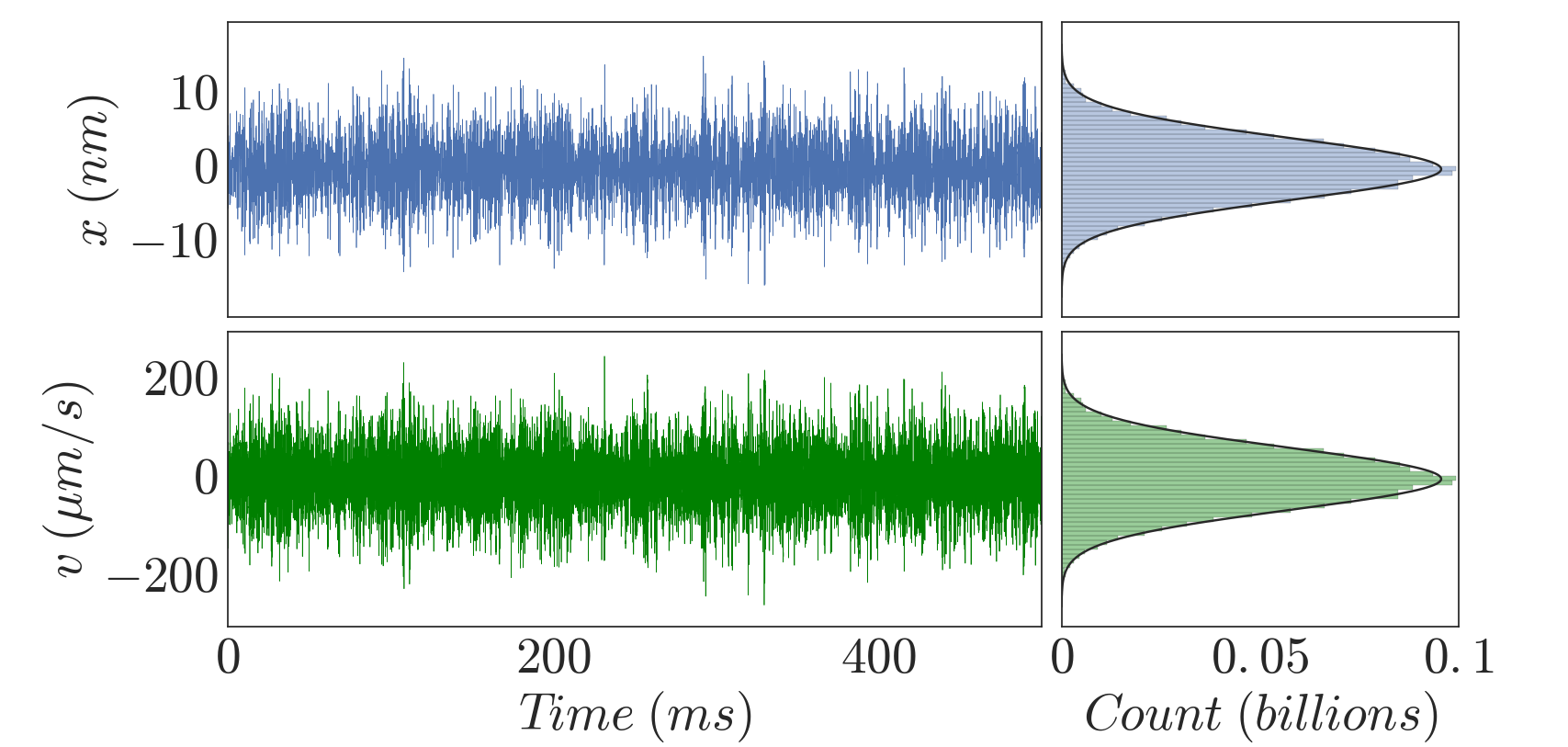}\caption{Sample paths (left) and histograms (right) of the position and velocity
of the Brownian harmonic oscillator obtained from exact path sampling,
Eq. (\ref{eq:path-sampling}). Parameters used are$m=1ng$, $k=225mg/s^{2}$,
$\gamma=\gamma_{c}/10$, $\gamma_{c}=\sqrt{4mk}$ and $T=275K$. The
path is sampled at $2^{16}Hz$. \label{fig:PositionsVelocity-1}}
}
\end{figure}

\textcolor{black}{In this appendix, we provide explicit expressions
for $e^{-\boldsymbol{\lambda}\Delta t}$ and $\sqrt{\boldsymbol{\Sigma}}$,
needed for the path sampling of a bivariate Ornstein-Uhlenbeck process
in section \ref{sec:Brownian-harmonic-oscillator}. From the Cayley-Hamilton
theorem, $e^{-\boldsymbol{\lambda}\Delta t}$ is obtained in terms
of the identity matrix and $\boldsymbol{\lambda}$ as shown in Eq.(\ref{eq:cayHam}).
The coefficients are\begin{subequations}
\begin{eqnarray}
\Lambda_{1} & = & \exp(\tfrac{-\Delta t}{2\tau})\left[\cos(\omega\Delta t)+\frac{1}{2\omega\tau}\sin(\omega\Delta t)\right],\\
\Lambda_{2} & = & \exp(\tfrac{-\Delta t}{2\tau})\left[-\frac{1}{\omega}\sin(\omega\Delta t)\right].
\end{eqnarray}
\end{subequations} Here $\omega=\sqrt{\omega_{0}^{2}-1/(2\tau)^{2}}$
is the frequency of the damped oscillator. $\sqrt{\boldsymbol{\Sigma}}$
is obtained from its Cholesky factorization into a lower triangular
matrix and its transpose, as shown in Eq.(\ref{eq:choleskyF}). The
elements of the Cholesky factor are\begin{subequations}
\begin{align}
s_{1} & =\bigg[\frac{k_{B}T}{k}-\frac{k_{B}T}{k}\exp(\tfrac{-\Delta t}{\tau})\Big(\frac{k}{m\omega^{2}}\sin^{2}(\omega\Delta t)+\big[\cos(\omega\Delta t)+\tfrac{1}{2\omega\tau}\sin(\omega\Delta t)\big]^{2}\Big)\bigg]^{1/2},\\
s_{2} & =\frac{1}{s_{1}}\dfrac{k_{B}T}{\gamma\omega^{2}\tau^{2}}\exp(\tfrac{-\Delta t}{\tau})\sin^{2}(\omega\Delta t),\\
s_{3} & =\bigg[\frac{k_{B}T}{m}-\frac{k_{B}T}{m}\exp(\tfrac{-\Delta t}{\tau})\Big(\frac{k}{m\omega^{2}}\sin^{2}(\omega\Delta t)+\big[\cos(\omega\Delta t)-\tfrac{1}{2\omega\tau}\sin(\omega\Delta t)\big]^{2}\Big)-s_{2}^{2}\bigg]^{1/2}.
\end{align}
\end{subequations}We use the above to elucidate the results in Eq.(\ref{eq:path-sampling})
in order to obtain exactly sampled trajectories of the Brownian harmonic
oscillator. In Fig.(\ref{fig:PositionsVelocity-1}) we show a typical
sample path of the positions and velocities and their corresponding
histograms. The histograms of the positions and the velocities clearly
show that the distributions are normal. }


\begin{thebibliography}{10}

\bibitem{doob1942brownian}
J.~L. Doob.
\newblock The {B}rownian movement and stochastic equations.
\newblock {\em Ann. Math.}, pages 351--369, 1942.

\bibitem{doob1953stochastic}
J.~L. Doob.
\newblock {\em Stochastic processes}, volume~7.
\newblock Wiley New York, 1953.

\bibitem{van1992stochastic}
N.~G. van Kampen.
\newblock {\em Stochastic processes in physics and chemistry}.
\newblock Elsevier, 1992.

\bibitem{gardiner1985handbook}
C.~W. Gardiner.
\newblock {\em Handbook of stochastic methods}.
\newblock Springer Berlin, 1985.

\bibitem{mackay2003information}
D.~J.~C. MacKay.
\newblock {\em Information theory, inference and learning algorithms}.
\newblock Cambridge University Press, 2003.

\bibitem{rasmussen2006gaussian}
C.~E. Rasmussen and C.~K.~I. Williams.
\newblock {\em Gaussian processes for machine learning}, volume~1.
\newblock MIT press Cambridge, 2006.

\bibitem{lindgren2011explicit}
F.~Lindgren, H.~Rue, and J.~Lindstr{\"o}m.
\newblock An explicit link between {G}aussian fields and {G}aussian {M}arkov
  random fields: the stochastic partial differential equation approach.
\newblock {\em J. Royal Stat. Soc.: Series B (Stat. Meth.)}, 73(4):423--498,
  2011.

\bibitem{smola2000sparse}
A.~J. Smola and B.~Sch{\"o}lkopf.
\newblock Sparse greedy matrix approximation for machine learning.
\newblock pages 911--918. Morgan Kaufmann, San Francisco, 2000.

\bibitem{williams2001using}
C.~K.~I. Williams and M.~Seeger.
\newblock Using the {N}ystr{\"o}m method to speed up kernel machines.
\newblock In {\em Advances in neural information processing systems}, pages
  682--688. MIT press, 2001.

\bibitem{csato2002sparse}
L.~Csat{\'o} and M.~Opper.
\newblock Sparse on-line {G}aussian processes.
\newblock {\em Neur. Comp.}, 14(3):641--668, 2002.

\bibitem{quinonero2005unifying}
J.~Qui{\~n}onero-Candela and C.~E. Rasmussen.
\newblock A unifying view of sparse approximate {G}aussian process regression.
\newblock {\em J. Mach. Learn. Res.}, 6:1939--1959, 2005.

\bibitem{kavcic2000matrices}
A.~Kavcic and J.~M.~F. Moura.
\newblock Matrices with banded inverses: {I}nversion algorithms and
  factorization of {G}auss-{M}arkov processes.
\newblock {\em IEEE Trans. Inf. Theory}, 46(4):1495--1509, 2000.

\bibitem{jeffreys1998theory}
H.~Jeffreys.
\newblock {\em The theory of probability}.
\newblock Oxford University Press, Oxford, 1939.

\bibitem{jaynes2003probability}
E.~T. Jaynes.
\newblock {\em Probability theory: The logic of science}.
\newblock Cambridge University Press, 2003.

\bibitem{sivia2006data}
D.~Sivia and J.~Skilling.
\newblock {\em Data analysis: a {B}ayesian tutorial}.
\newblock Oxford University Press, Oxford, 2006.

\bibitem{bera2017fast}
S.~Bera, S.~Paul, R.~Singh, D.~Ghosh, A.~Kundu, A.~Banerjee, and R.~Adhikari.
\newblock Fast {B}ayesian inference of optical trap stiffness and particle
  diffusion.
\newblock {\em Sci. Rep.}, 7:41638, 2017.

\bibitem{kashyap1977bayesian}
R.~Kashyap.
\newblock A bayesian comparison of different classes of dynamic models using
  empirical data.
\newblock {\em IEEE Trans. Automat. Cont.}, 22(5):715--727, 1977.

\bibitem{mackay1992bayesian}
D.~J.~C. MacKay.
\newblock Bayesian interpolation.
\newblock {\em Neur. Computn.}, 4:415--447, 1992.

\bibitem{zellner1984}
A.~Zellner.
\newblock {\em Basic Issues in Econometrics}.
\newblock University of Chicago Press, Chicago, 1984.

\bibitem{gregory2005bayesian}
P.~Gregory.
\newblock {\em Bayesian Logical Data Analysis for the Physical Sciences}.
\newblock Cambridge University Press, 2005.

\bibitem{uhlenbeck1930theory}
G.~E. Uhlenbeck and L.~S. Ornstein.
\newblock On the theory of the {B}rownian motion.
\newblock {\em Phys. Rev.}, 36(5):823, 1930.

\bibitem{parker2012sampling}
A.~Parker and C.~Fox.
\newblock Sampling {G}aussian distributions in {K}rylov spaces with conjugate
  gradients.
\newblock {\em SIAM J. Sci. Comput.}, 34(3):B312--B334, 2012.

\bibitem{chandrasekhar1943stochastic}
S.~Chandrasekhar.
\newblock Stochastic problems in physics and astronomy.
\newblock {\em Rev. Mod. Phys.}, 15:1--89, 1943.

\bibitem{einstein1905theory}
A.~Einstein.
\newblock The theory of the {B}rownian movement.
\newblock {\em Ann. Phys. (Berlin)}, 322:549, 1905.

\bibitem{kubo1966fluctuation}
R.~Kubo.
\newblock The fluctuation-dissipation theorem.
\newblock {\em Rep. Prog. Phys.}, 29(1):255, 1966.

\bibitem{foreman2016corner}
D.~Foreman-Mackey.
\newblock corner.py: Scatterplot matrices in {P}ython.
\newblock {\em The Journal of Open Source Software}, 1(2):1--2, 2016.

\bibitem{burg1967maximum}
J.~P. Burg.
\newblock {\em Maximum entropy spectral analysis.}
\newblock PhD thesis, Stanford University, 1975.

\bibitem{berg2004power}
K.~Berg-S{\o}rensen and H.~Flyvbjerg.
\newblock Power spectrum analysis for optical tweezers.
\newblock {\em Rev. Sci. Inst.}, 75(3):594--612, 2004.

\bibitem{tassieri2012microrheology}
M.~Tassieri, R.~Evans, R.~L. Warren, N.~J. Bailey, and J.~M. Cooper.
\newblock Microrheology with optical tweezers: data analysis.
\newblock {\em New J. Phys.}, 14(11):115032, 2012.

\bibitem{Bera:16}
S.~Bera, A.~Kumar, S.~Sil, T.~K. Saha, T.~Saha, and A.~Banerjee.
\newblock Simultaneous measurement of mass and rotation of trapped absorbing
  particles in air.
\newblock {\em Opt. Lett.}, 41(18):4356--4359, 2016.

\bibitem{gardiner1984adiabatic}
C.~W. Gardiner.
\newblock Adiabatic elimination in stochastic systems. {I}. {F}ormulation of
  methods and application to few-variable systems.
\newblock {\em Phys. Rev. A}, 29(5):2814--2822, 1984.

\end{thebibliography}
\end{document}